\documentclass[twocolumn,preprintnumbers,amsmath,amssymb,pre,nofootinbib]{revtex4}
\usepackage{multirow}
\usepackage{graphics}
\usepackage{array}
\newcommand{\ket}[1]{\ensuremath{\left| #1 \right\rangle}}
\newcommand{\br}[1]{\ensuremath{\left\langle #1 \right.}}
\newcommand{\bra}[1]{\ensuremath{\left. \br{#1} \right|}}
\newcommand{\bk}[2]{\br{{#1}}\ket{{#2}}}
\newcommand{\kb}[2]{\ket{{#1}} \! \bra{{#2}}}
\newcommand{\proj}[1]{\kb{{#1}}{{#1}}}

\newcommand{\modu}[1]{\ensuremath{\left| {#1} \right|}}
\newcommand{\magn}[1]{\ensuremath{\modu{#1}^2}}
\newcommand{\born}[2]{\ensuremath{\magn{\bk{{#1}}{{#2}}}}}

\newcommand{\pontic}{$\Psi$-ontic }
\newcommand{\peptic}{$\Psi$-epistemic }

\newcommand{\dofep}[2]{\Omega \left[ #1,#2 \right ]}

\newcommand{\dthird}{\frac{1}{3}}
\newcommand{\dtwothird}{\frac{2}{3}}
\newcommand{\dhalf}{\ensuremath{\frac{1}{2}}}

\newcommand{\half}{\ensuremath{\frac{1}{2}}}

\begin{document}
\title{How statistical are quantum states?}
\author{O.~J.~E.~Maroney\footnote{email:owen.maroney@philosophy.ox.ac.uk}}
\affiliation{Faculty of Philosophy, University of Oxford, 10 Merton Street, Oxford, OX1 4JJ, UK\footnote{Mailing address:  Wolfson College, Linton Road, Oxford, OX2 6UD, UK}}
\date{\today}

\begin{abstract}
A novel no-go theorem is presented which sets a bound upon the extent to which `\peptic' interpretations of quantum theory are able to explain the overlap between non-orthogonal quantum states in terms of an experimenter's ignorance of an underlying state of reality.  The theorem applies to any Hilbert space of dimension greater than two.  In the limit of large Hilbert spaces, no more than half of the overlap between quantum states can be accounted for.  Unlike other recent no-go theorems no additional assumptions, such as forms of locality, invasiveness, or non-contextuality, are required.
\end{abstract}

\maketitle

One of the most remarkable features of a quantum system is that of non-orthogonality: it is possible to prepare two different pure quantum states which cannot be perfectly discriminated by a single ideal measurement.  No such property exists in classical systems, where two different classical states can always by perfectly discriminated by a suitably ideal measurement.

Statistical interpretations of the quantum state (see \cite{Ballentine1970,Spekkens2007,HS2007,BRS2012} and references therein) present a natural way to understand this feature of non-orthogonality. Suppose that a `quantum state' is only a statistical state, associated with a probability distribution over some underlying physical states. Even an ideal quantum preparation procedure cannot uniquely fix this underlying state of the system but instead prepares one of a number of possible such states, with a well defined probability.  In such an interpretation, when two quantum states are non-orthogonal, the associated probability distributions will overlap: the preparation procedures each have a non-zero probability of preparing the same physical states, and this is why they cannot be perfectly distinguished.  This has been called a \peptic interpretation\cite{Spekkens2007,HS2007}, as the idea is that the wavefunction does not itself represent a physical state, but instead represents an epistemic uncertainty about the physical state.

A number of recent no-go theorems\cite{PBR2012,CR2011,SF2012,Hardy2012,PPM2013,ABCL2013} have explored the consequences of attempting to construct such \peptic interpretations of quantum theory.  This paper contributes a new no-go theorem, demonstrating that any attempt to explain non-orthogonality entirely in terms of an epistemic overlap is impossible: there are no `maximally epistemic' interpretations of quantum theory.  Using an empirically testable measure, it is shown that in the limit of large Hilbert spaces, no more than half of the overlap between quantum states could be explained in terms of an overlap of probability distributions.  Unlike the other recent results, the theorem presented here makes no additional assumptions beyond the definition of a \peptic theory.  This means that this theorem is the only such theorem to apply, for example, to the type of constructive \peptic theories presented in \cite{LJBR2012,ABCL2013}.

\textit{Ontological models and $\Psi$-epistemic theories.-}
Two principal assumptions are involved in analysing whether a theory is \peptic:
\begin{enumerate}
\item After a physical procedure prepares a quantum state, $\ket{\psi}$, the system is actually in a physical state $\lambda$. This physical state is not necessarily identified with the quantum state;
\item The probabilities of getting the results of a measurement procedure on the system are wholly determined by the physical state $\lambda$ and the measurement procedure.  The preparation procedure only influences the measurement outcomes indirectly, through the possible physical states prepared.
\end{enumerate}
These assumptions are developed into the \textit{ontological models} formalism in\cite{Spekkens2005a,HS2007,HR2007,Rudolph2006,AHR2008}, where the physical state $\lambda$, is referred to as an \textit{ontic} state.
\begin{itemize}
\item A preparation procedure, which prepares the quantum state $\ket{\psi}$, will prepare some ontic state $\lambda$ with probability $\mu_{\psi}(\lambda)$:
\begin{equation}
\mu_{\psi}(\lambda) \ge 0, \quad
\int \mu_{\psi}(\lambda) d\lambda=1
\end{equation}
\item A measurement, $M$, has a number of possible outcomes $\{Q\}$, and a probability $\xi_M(Q|\lambda)$ of obtaining a particular outcome $Q$, given the ontic state $\lambda$:

    \begin{equation}
\xi_M(Q|\lambda) \ge 0, \quad
\sum_Q \xi_M(Q| \lambda) =1
\end{equation}
\item An ontological model will reproduce the results of quantum theory if, and only if:
\begin{equation}
\int \mu_\psi(\lambda)\xi_M(Q|\lambda)d\lambda
=
\magn{\bk{Q}{\psi}}
 \end{equation}
\end{itemize}
Additional structure is required to represent unitary operations, but this is not needed for this proof.

Some immediate consequences of this formalism will be useful.  The set $\Lambda_{\phi}=\{\lambda : \mu_{\phi}(\lambda)>0\}$ is the set of all possible ontic states which may occur when preparing the quantum state $\ket{\phi}$.

If one of the outcomes of the measurement $M$ is represented by the projection onto $\ket{\phi}$, then
$
\int \mu_{\phi}(\lambda)\xi_M(\phi|\lambda)d\lambda
=
\magn{\bk{\phi}{\phi}}=1
$.  It follows that
\begin{equation}
\forall \lambda \in \Lambda_{\phi} \;\;\; \xi_M(\phi|\lambda)=1
\end{equation}

If $\ket{\varphi}$ is orthogonal to $\ket{\phi}$ then
$
\int \mu_{\varphi}(\lambda)\xi_M(\phi|\lambda)d\lambda
=
\magn{\bk{\phi}{\varphi}}=0
$.  It follows that
\begin{equation}
\forall \lambda \in \Lambda_{\varphi} \;\;\; \xi_M(\phi|\lambda)=0
\end{equation}
which implies $\Lambda_{\varphi} \cap \Lambda_{\phi}=\emptyset$.  (Sets of measure zero can be ignored.)

If the quantum two states $\ket{\psi}$ and $\ket{\phi}$ are non-orthogonal, then:
\begin{equation}
\int \mu_{\psi}(\lambda)\xi_M(\phi|\lambda)d\lambda
=
\magn{\bk{\phi}{\psi}} \neq 0
 \end{equation}
This allows the possibility that $\Lambda_{\psi} \cap \Lambda_{\phi}$ is not empty.

A \peptic theory is defined\cite{HS2007} to be one in which there exist at least two distinct non-orthogonal quantum states, $\ket{\psi}$ and $\ket{\phi}$, for which $\Lambda_{\psi} \cap \Lambda_{\phi} \neq \emptyset$.  In such a theory, the fact that the quantum states $\ket{\psi}$ and $\ket{\phi}$ cannot be discriminated by a single ideal measurement is, at least partially, explained by the fact that there is some $\lambda \in \Lambda_{\psi} \cap \Lambda_{\phi}$.  As all the results of any measurement $M$ are determined by $\xi_M(Q|\lambda)$, and both $\ket{\psi}$ and $\ket{\phi}$ can prepare ontic states in $\Lambda_{\psi} \cap \Lambda_{\phi}$, it follows that there is no possible measurement outcome which can occur with zero probability for $\ket{\psi}$ but with certainty for $\ket{\phi}$ (and vice versa).

The term $\int_{\Lambda_{\phi}} \mu_{\psi}(\lambda) d\lambda$ gives a measure of the epistemic overlap between the quantum states $\ket{\psi}$ and $\ket{\phi}$.  This is bound by the measure of the quantum state overlap. As $\forall \lambda \in \Lambda_{\phi}$,  $\xi_M(\phi|\lambda)=1$
\begin{align}
\int_{\Lambda_{\phi}} \mu_{\psi}(\lambda) d\lambda& =\int_{\Lambda_{\phi}} \mu_{\psi}(\lambda)\xi_M(\phi|\lambda)d\lambda \nonumber \\
 & \le \int \mu_{\psi}(\lambda)\xi_M(\phi|\lambda)d\lambda=\magn{\bk{\phi}{\psi}}
\end{align}

The degree of epistemic overlap, $\dofep{\phi}{\psi}$, between two states can now be defined via:
\begin{equation}
\int_{\Lambda_{\phi}} \mu_{\psi}(\lambda)d\lambda
=\dofep{\phi}{\psi}\magn{\bk{\phi}{\psi}}
 \end{equation}
By definition, $\dofep{\psi}{\psi}=1$.  Where $\bk{\phi}{\psi}=0$, $\dofep{\phi}{\psi}$ may take any value.

A \textit{maximally \peptic} theory would have $\dofep{\phi}{\psi}=1$ for all pairs of quantum states.  The quantum state overlap $\magn{\bk{\phi}{\psi}}$ would be accounted for entirely in terms of the overlap in the ontic states $\Lambda_{\psi} \cap \Lambda_{\phi}$.

By contrast, a \textit{maximally \pontic} theory\footnote{In much of the existing literature this is simply referred to as a \pontic theory.} is one in which $\Lambda_{\psi} \cap \Lambda_{\phi}=\emptyset$ for all pairs of distinct quantum states, and so $\dofep{\phi}{\psi}=0$ .  In such a theory, no two distinct quantum state preparations $\ket{\psi}$ and $\ket{\phi}$ could possibly produce the same ontic state $\lambda$.  For a maximally \pontic theory, therefore, given the ontic state $\lambda$ it is always possible to identify the quantum state $\ket{\psi}$.  The de Broglie-Bohm\cite{BH93,Hol93}, Everett\cite{BKSW2010} and CSL\cite{BG2003} interpretations of quantum theory are all examples of maximally \pontic theories.

It will now be shown that, in Hilbert spaces of dimension greater than two, there are no models of quantum theory for which $\dofep{\phi}{\psi}=1$ for all states: there are no maximally \peptic interpretations.  Making a symmetrical assumption that $\dofep{\phi}{\psi}$ is a constant, $\Omega$, between all pairs of distinct states it will be shown that $\Omega \le \frac{9}{10}$ for a 3 dimensional Hilbert space, falling to $\Omega \le \half$ in the limit of large Hilbert spaces.  No more than 50\% of the quantum state overlap can be accounted for by an epistemic overlap of probability distributions.

\textit{Three dimensional Hilbert space.-}
Consider the following states of a three dimensional Hilbert space:
\[
\begin{array}{c}
\ket{a},  \; \; \ket{b},  \; \; \ket{c}\\
\ket{p}=\frac{1}{\sqrt{3}}\left(\ket{a}+\ket{b}+\ket{c}\right), \;
\ket{m}=\frac{1}{\sqrt{3}}\left(\ket{a}+\ket{b}-\ket{c}\right)\\
\ket{a_+}=\frac{1}{\sqrt{2}}\left(\ket{a}+\ket{c}\right), \;
\ket{a_-}=\frac{1}{\sqrt{2}}\left(\ket{a}-\ket{c}\right)\\
\ket{b_+}=\frac{1}{\sqrt{2}}\left(\ket{b}+\ket{c}\right), \;
\ket{b_-}=\frac{1}{\sqrt{2}}\left(\ket{b}-\ket{c}\right)\\
\end{array}
\]
where $\ket{a}, \ket{b}, \ket{c}$ form an orthonormal basis, with the three measurements:
\[
\begin{array}{cl}
M_1 : & \proj{a_+},\proj{a_-},\proj{b}\\
M_2 : & \proj{b_+},\proj{b_-},\proj{a}\\
M_3 : & \proj{a},\proj{b},\proj{c} \\
\end{array}
\]

(These are the states and measurements used in the Clifton-Stairs proof of quantum contextuality\cite{Stairs1983,Clifton1993}.  For further connections to quantum contextuality, see\cite{LM2013}.)

\begin{table}[htb]
\centering
\textbf{(a)}
\[
\begin{array}{c|c c c}\hline \hline
  &\multicolumn{3}{c}{M_1}\\
    & \proj{b} & \proj{a_+} & \proj{a_-} \\ \hline
\ket{a}  & 0 & \dhalf & \dhalf  \\
\ket{b}  & 1 & 0 & 0   \\
\ket{c}  & 0 & \dhalf & \dhalf \\
\ket{p}  & \dthird & \dtwothird & 0  \\
\ket{m}  & \dthird & 0 & \dtwothird \\ \hline \hline
\end{array}
\]
\textbf{(b)}
\[
\begin{array}{c|c c c} \hline \hline
 & \multicolumn{3}{c}{M_2} \\
    & \proj{a} & \proj{b_+} & \proj{b_-} \\ \hline
\ket{a} &  1 & 0 & 0 \\
\ket{b} &  0 & \dhalf & \dhalf \\
\ket{c} &  0 & \dhalf & \dhalf \\
\ket{p} &  \dthird & \dtwothird & 0\\
\ket{m} &  \dthird & 0 & \dtwothird \\ \hline \hline
\end{array}
\]
\textbf{(c)}
\[
\begin{array}{c|c c c} \hline \hline
        & \multicolumn{3}{c}{M_3} \\
        & \proj{a} & \proj{b} & \proj{c} \\ \hline
\ket{a} &  1 & 0 & 0 \\
\ket{b} &  0 & 1 & 0  \\
\ket{c} &  0 & 0 & 1 \\
\ket{p} & \dthird & \dthird & \dthird \\
\ket{m} & \dthird & \dthird & \dthird \\ \hline \hline
\end{array}
\]
\caption{Measurement outcomes}
\label{t:m1}
\end{table}

The results are shown in Table \ref{t:m1}.  From $M_1$ it can be seen that
\begin{equation}\begin{array}{l}
\forall \lambda \in \Lambda_{a}, \;\;\;  \xi_{M_1}(b|\lambda)=0 \\
\forall \lambda \in \Lambda_{p}, \;\;\;  \xi_{M_1}(a_-|\lambda)=0 \\
\forall \lambda \in \Lambda_{m}, \;\;\;  \xi_{M_1}(a_+|\lambda)=0
\end{array}
\end{equation}
For any given $\lambda$, it must be that $\xi_{M_1}(b|\lambda)+\xi_{M_1}(a_-|\lambda)+\xi_{M_1}(a_+|\lambda)=1$, so this means:
\begin{equation}
\Lambda_{a} \cap \Lambda_{p} \cap \Lambda_{m}=\emptyset
\end{equation}
Similarly
\begin{equation}
\Lambda_{c} \cap \Lambda_{p} \cap \Lambda_{m}=\emptyset
\end{equation}
$M_2$ adds
\begin{equation}
\Lambda_{b} \cap \Lambda_{p} \cap \Lambda_{m}=\emptyset
\end{equation}
leading to
\begin{equation}
\left(\Lambda_{a}\cup \Lambda_{b}\cup\Lambda_{c}\right)\cap \Lambda_{p} \cap \Lambda_{m}=\emptyset
\end{equation}
If there is any overlap in $\Lambda_{p} \cap \Lambda_{m}$ then it must lie outside $\Lambda_{a}\cup \Lambda_{b}\cup\Lambda_{c}$.

Now $M_3$ gives
\begin{equation}
\int_{\Lambda_{a}} \mu_{p}(\lambda)d\lambda =\frac{\dofep{a}{p}}{3}
 \end{equation}
and similarly for $\ket{b}$ and $\ket{c}$.  As $\Lambda_{b} \cap \Lambda_{a}=\emptyset$ etc.
\begin{equation}
\int_{\left(\Lambda_{a}\cup \Lambda_{b} \cup \Lambda_{c}\right)} \mu_{p}(\lambda) d\lambda =\frac{\left(\dofep{a}{p}+\dofep{b}{p}+\dofep{c}{p}\right)}{3}
 \end{equation}
It follows that:
\begin{equation}
\int_{\Lambda_{m}}\mu_{p}(\lambda)  d\lambda \le 1-\frac{\left(\dofep{a}{p}+\dofep{b}{p}+\dofep{c}{p}\right)}{3}
 \end{equation}
 but
 \begin{equation}
 \int_{\Lambda_{m}}  \mu_{p}(\lambda)d\lambda=\frac{\dofep{m}{p}}{9}
 \end{equation} giving
 \begin{equation}
 \frac{\dofep{m}{p}}{9} \le 1-\frac{\left(\dofep{a}{p}+\dofep{b}{p}+\dofep{c}{p}\right)}{3}
 \end{equation}
If the ontic model has maximal epistemic overlaps with the $\ket{a},\ket{b},\ket{c}$ states, then $\dofep{a}{p}=\dofep{b}{p}=\dofep{c}{p}=1$, but this requires $\dofep{m}{p}=0$ while $\magn{\bk{m}{p}}=\frac{1}{9}$.  This proves that a maximally \peptic theory is not possible.

Reducing the degree of epistemicness in the $\ket{a},\ket{b},\ket{c}$ basis will allow some epistemic overlap between $\ket{p}$ and $\ket{m}$.  For example, keeping $\dofep{a}{p}=\dofep{b}{p}=1$ it is possible to reach $\dofep{m}{p}=1$ with $\dofep{c}{p}=\frac{2}{3}$.  More symmetrically, $\dofep{m}{p}=1$ is possible with $\dofep{a}{p}=\dofep{b}{p}=\dofep{c}{p}=\frac{8}{9}$.

A basis independent measure of how epistemic a theory can be and still reproduce quantum statistics is given by letting $\dofep{\phi}{\psi}=\Omega$ be a constant for all non-orthogonal states.  This gives:
\begin{equation}
\Omega \le \frac{9}{10}
\end{equation}

\textit{Higher dimensional Hilbert spaces.-}
The above results can be generalised to Hilbert spaces of dimension $d>3$. In the limit of large $d$,  $\Omega \le \half$.

Take the basis states
\[
\ket{a_1},\ket{a_2},\ldots \ket{a_d}
\]
the superpositions
\[
\begin{array}{l}
\ket{a_{i+}}=\frac{1}{\sqrt{2}}\left(\ket{a_i}+\ket{a_d}\right)\\
\ket{a_{i-}}=\frac{1}{\sqrt{2}}\left(\ket{a_i}-\ket{a_d}\right)\\
\ket{p_d}=\frac{1}{\sqrt{d}}\sum_{i=1,d} \ket{a_i}\\
\ket{m_d}=\frac{1}{\sqrt{d}}\left(\sum_{i=1,d-1}\ket{a_i}\right)-\frac{1}{\sqrt{d}}\ket{a_d}
\end{array}
\]
and the measurements
\[
\begin{array}{cl}
M_i : & \proj{a_1},\ldots,\proj{a_{i-1}}, \\
    & \qquad \proj{a_{i+1}},\ldots,\proj{a_{d-1}},\\
    & \qquad \qquad \proj{a_{i+}},\proj{a_{i-}}\\
M_d : & \proj{a_1},\dots,\proj{a_d}
\end{array}
\]
Each measurement $M_i$ now shows:
\begin{equation}
\begin{array}{l}
\Lambda_{a_i} \cap \Lambda_{p_d} \cap \Lambda_{m_d}=\emptyset \\
\Lambda_{a_d} \cap \Lambda_{p_d} \cap \Lambda_{m_d}=\emptyset
\end{array}
\end{equation}
leading to
\begin{equation}
\left(\cup_i \Lambda_{a_i}\right)\cap \Lambda_{p_d} \cap \Lambda_{m_d}=\emptyset
\end{equation}
From the $M_d$ measurement:
\begin{equation}
\int_{\left(\cup_i \Lambda_{a_i}\right)} \mu_{p_d}(\lambda) d\lambda  =\sum_i \dofep{a_i}{p_d}\magn{\bk{a_i}{p_d}}=\frac{1}{d}\sum_i \dofep{a_i}{p_d}
 \end{equation}
so
\begin{equation}
\int_{\Lambda_{m_d}} \mu_{p_d}(\lambda) d\lambda \le 1-\frac{1}{d}\sum_i \dofep{a_i}{p_d}
 \end{equation}
 giving
  \begin{equation}
 \dofep{m_d}{p_d}\left(1-\frac{2}{d}\right)^2 \le 1-\frac{1}{d}\sum_i \dofep{a_i}{p_d}
 \end{equation}
 Once again, $\dofep{a_i}{p_d}=1$ would require $\dofep{m_d}{p_d}=0$, even though $\magn{\bk{m_d}{p_d}}=\left(1-\frac{2}{d}\right)^2 \rightarrow 1$.  For a theory to be maximally \peptic in a specific basis, then as $d \rightarrow \infty$ it must become maximally \pontic between at least some states arbitrarily close to each other.  To obtain a full overlap $\dofep{m_d}{p_d}=1$ by symmetrically reducing the overlap with the $\ket{a_i}$ basis would require $\dofep{a_i}{p_d}=\frac{4}{d^2}\left(d-1\right) \rightarrow 0$.

The basis independent measure of how epistemic this allows the theory overall sets $\dofep{\phi}{\psi}=\Omega(d)$
as before giving
\begin{equation}
\Omega(d) \le \frac{d^2}{2d^2-4d+4}
\end{equation}
In the limit $d$ gets large, $\Omega(d) \le \half$.

\textit{Experimental Noise.-}
The calculations depend upon obtaining zero probability of obtaining a set of outcomes for particular measurements.  In any real experiment, there will be some noise, giving a value $\epsilon>0$.  In such cases the results calculated above become unstable.

Following \cite{PBR2012}, the measure of overlap in the ontic state distribution which can be compared to experimental data in the presence of noise is the quantum trace norm distance, rather than the quantum state overlap:
\begin{equation}
\delta_Q(\rho,\sigma)=\half \textrm{Tr}\magn{\rho-\sigma}
\end{equation}
This has an operational interpretation: if one of two quantum states, $\rho$ or $\sigma$ is prepared with equal probability, then using an optimal quantum measurement, the probability of correctly identifying which is given by
\begin{equation}
P_Q=\half(1+\delta_Q(\rho,\sigma))
\end{equation}
This should be compared to the classical trace norm distance
\begin{equation}
\delta_C(p,q)=\half \int \magn{p(x)-q(x)}dx
\end{equation}
which also has an operational interpretation: if a random variable $x$ is prepared by one of two methods, producing classical distributions $p(x)$ and $q(x)$ respectively, the optimal probability of guessing which method was used is given by  
\begin{equation}
P_C=\half(1+\delta_C(p,q))
\end{equation}
In any ontological model for quantum theory, the measuring device is presented with an ontic state $\lambda$, prepared with the probability distribution $\mu_\phi(\lambda)$ or $\mu_\psi(\lambda)$.  As this is the most information the optimal measuring device has available, the probability of guessing which quantum state $\phi$ or $\psi$ was prepared, based only on the outputs of the measuring device, cannot be greater than the probability of guessing based upon the ontic state itself, so $P_Q \le P_C$.

The quantum trace distance for pure states is
\begin{equation}
\delta_Q(\phi,\psi)=\sqrt{1-\born{\phi}{\psi}}
\end{equation}
while the classical trace norm distance can be expressed as
\begin{equation}
\delta_C(\phi,\psi)=1-\int \min[\mu_\phi(\lambda),\mu_\phi(\lambda)]d\lambda
\end{equation}
leading to
\begin{equation}
\int \min[\mu_\phi(\lambda),\mu_\phi(\lambda)]d\lambda \le 1-\sqrt{1-\born{\phi}{\psi}}
\end{equation}
In a maximally \pontic theory, the left hand side is zero, while in a maximally \peptic theory the equality should be reached.  A routine but lengthy calculation shows that for the experimental arrangements considered here, the inequality is strict between some pairs of states in all Hilbert spaces of dimension $d>14$.  
\begin{equation}
\frac
{\int \min[\mu_\phi(\lambda),\mu_\phi(\lambda)]d\lambda}
{1-\sqrt{1-\born{\phi}{\psi}}} < 1 
\end{equation}
A discussion of this measure, as well as an experimental arrangement better suited to comparing the trace norm distances which extends the result to all Hilbert spaces of dimension $d>2$ will be discussed in the forthcoming \cite{BCLM2013}.

\textit{Conclusions.-} There exist constructive examples of \peptic models, which can reproduce all the statistics of quantum theory (such as the LJBR model\cite{LJBR2012} and the ABCL model\cite{ABCL2013}).  It is therefore not the intention of this article to attempt to construct no-go theorems which aim to show that \peptic theories are impossible simpliciter.  Existing no-go theorems have effectively sought to prove $\dofep{\phi}{\psi}=0$, but to do so must assume additional properties, such as forms of locality\cite{CR2011}, factorisability of product states\cite{PBR2012} or non-invasiveness\cite{Hardy2012}.  This should not downplay the significance of these theorems: any attempt to construct \peptic theories must necessarily reject these properties, in much the same way that Bell's theorem\cite{Bell1964} requires realist interpretations of quantum theory to reject local causality.  They help to map out the space of possible theories which can account for quantum phenomena.

The no-go theorem presented here contributes to this task.  It does not make additional assumptions so applies even to theories that reject all of these additional assumptions, such as the LJBR and ABCL models.  These models demonstrate that $\dofep{\phi}{\psi}>0$ is indeed possible.  This no-go theorem shows that $\dofep{\phi}{\psi}<1$, and this result follows directly from the definition of a \peptic model, even if all other assumptions are dropped.

{\center

\textbf{Acknowledgements}

}

I would like to thank Chris Timpson, Richard George, Harvey Brown, George Knee, Andrew Briggs, Erik Gauger, Simon Benjamin and Jon Barrett for helpful discussions and comments.  This research is supported by the John Templeton Foundation.

\bibliographystyle{apsrev}

\end{document}